\documentclass[review]{elsarticle}
\usepackage[latin9]{inputenc}
\usepackage{color}
\usepackage{array}
\usepackage{verbatim}
\usepackage{float}
\usepackage{url}
\usepackage{multirow}
\usepackage{algpseudocode}
\usepackage{amsmath}
\usepackage{amsthm}
\usepackage{graphicx}
\usepackage{amssymb}
\usepackage{amstext}
\usepackage[ruled,vlined,linesnumbered]{algorithm2e}
\usepackage{multirow}
\usepackage{booktabs}
\usepackage[x11names,dvipsnames,table]{xcolor} 
\usepackage{colortbl}

\definecolor{Mycolor1}{HTML}{9C1ED6}
\definecolor{Mycolor2}{HTML}{37AA84}

\makeatletter
\DeclareFontEncoding{LGR}{}{}

\ProvideTextCommand{\~}{LGR}[1]{\char126#1}

\newcommand{\lyxmathsym}[1]{\ifmmode\begingroup\def\b@ld{bold}
  \text{\ifx\math@version\b@ld\bfseries\fi#1}\endgroup\else#1\fi}

\providecommand{\tabularnewline}{\\}

  \theoremstyle{definition}

\usepackage[margin=1in]{geometry}
\newcommand{\subfour}[1]{\vspace*{3mm}{\noindent\bf #1}}

\@ifundefined{showcaptionsetup}{}{%
 \PassOptionsToPackage{caption=false}{subfig}}
\usepackage{subfig}
\makeatother

  \providecommand{\definitionname}{Definition}

\begin{document}
\begin{frontmatter}
%
\title{Adversarial Attacks on Speech Recognition Systems for Mission-Critical Applications: A Survey}

\author[deakin]{Ngoc Dung Huynh}
\ead{ndhuynh@deakin.edu.au}  
\author[deakin]{Mohamed Reda Bouadjenek}
\ead{reda.bouadjenek@deakin.edu.au}  
\author[deakin]{Imran Razzak}
\ead{imran.razzak@deakin.edu.au}
\author[deakin]{Kevin Lee}
\ead{kevin.lee@deakin.edu.au}
\author[deakin]{Chetan Arora}
\ead{chetan.arora@deakin.edu.au}
\author[deakin]{Ali Hassani}
\ead{ali.hassani@deakin.edu.au}
\author[deakin]{Arkady Zaslavsky}
\ead{arkady.zaslavsky@deakin.edu.au}

\address[deakin]{School of Information Technology, Deakin University, Waurn Ponds Campus, Geelong, VIC 3216, Australia}

\newtheorem{assumption}{\textbf{Assumption}}

\begin{abstract}
A Machine-Critical Application is a system that is fundamentally necessary to the success of specific and sensitive operations such as search and recovery, rescue, military, and emergency management actions.
Recent advances in Machine Learning, Natural Language Processing, voice recognition, and speech processing technologies have naturally allowed the development and deployment of speech-based conversational interfaces to interact with various machine-critical applications.
While these conversational interfaces have allowed users to give voice commands to carry out strategic and critical activities, their robustness to adversarial attacks remains uncertain and unclear.
%
Indeed, Adversarial Artificial Intelligence (AI) which refers to a set of techniques that attempt to fool machine learning models with deceptive data, is a growing threat in the AI and machine learning research community, in particular for machine-critical applications. 
The most common reason of adversarial attacks is to cause a malfunction in a machine learning model. 
An adversarial attack might entail presenting a model with inaccurate or fabricated samples as it's training data, or introducing maliciously designed data to deceive an already trained model.
While focusing on speech recognition for machine-critical applications, in this paper, we first review existing speech recognition techniques, then, we investigate the effectiveness of adversarial attacks and defenses against these systems, before outlining research challenges, defense recommendations, and future work. 
This paper is expected to serve researchers and practitioners as a reference to help them in understanding the challenges, position themselves and, ultimately, help them to improve existing models of speech recognition for mission-critical applications.

\end{abstract}


\end{frontmatter}

\section{Introduction}
A Mission-Critical Application is a system where a hazard can degrade or prevent the successful completion of an intended operation~\cite{Fowler2004}.
Examples of mission-critical applications include search and recovery, rescue, military, and emergency management actions.
Hence, if a mission-critical application is interrupted or faulty, there can be a direct  negative financial or life-threatening consequences. 
For instance, a bank can lose billions of dollars if its mission-critical system is defeated, or if a disruption occurs in an automatic ambulance locator system, it may impact a rescue operation causing loss of lives.
Overall, mission-critical applications depend on reliable and timely data, so that operators use the information to understand the current situation and to take the best actions on time.


Recent technological advancements in Artificial Intelligence (AI) have revived the concept of direct communication with computer systems. 
Most organizations are swallowed by technological hype in a quest for personalized, efficient, and convenient interactions with customers.  A simple but efficient solution is Conversational User Interfaces (CUI)~\cite{sanner:www21,Allen2001,Gao2018}. 
CUIs for mission-critical applications are the next great leap forward, with dual-way interaction between machine and human, helping end-users to solve their problems with voice support. 
Speech recognition is the core component of a voice-based conversational system, aiming to convert a speech from an audio form into a textual format for easier downstream processing. 
In the early days, Hidden Markov Model (HMM) \cite{Franzese:2019} was the primary tool for speech recognition. However, the development of this traditional method has saturated in terms of both latency and accuracy. 
With Deep Learning advancements~\cite{LeCun2015,NIPS2012_c399862d}, Deep Neural Networks (DNNs) replaced some of the standard system components. 
Deep learning has been applied to develop such conversational systems for mission-critical applications for instance to recognize the transcription of medical speech in healthcare \cite{Edwards:2017}.
In recent years, the trend is to design an end-to-end neural network and leverage a massive amount of data to improve the accuracy of speech conversational systems. 
In the end-to-end models, the modules of the traditional system (acoustic model, pronunciation model, and language model) are jointly optimized in a single system. 
Examples of end-to-end models include CTC-based models \cite{Graves:2006} and Attention-based models  \cite{Chan:2016, Bahdanau:2016, Chorowski:2015, Chorowski:2014, Cho:2014}.

\begin{figure}[t]
  \centering
  \includegraphics[width=.95\linewidth]{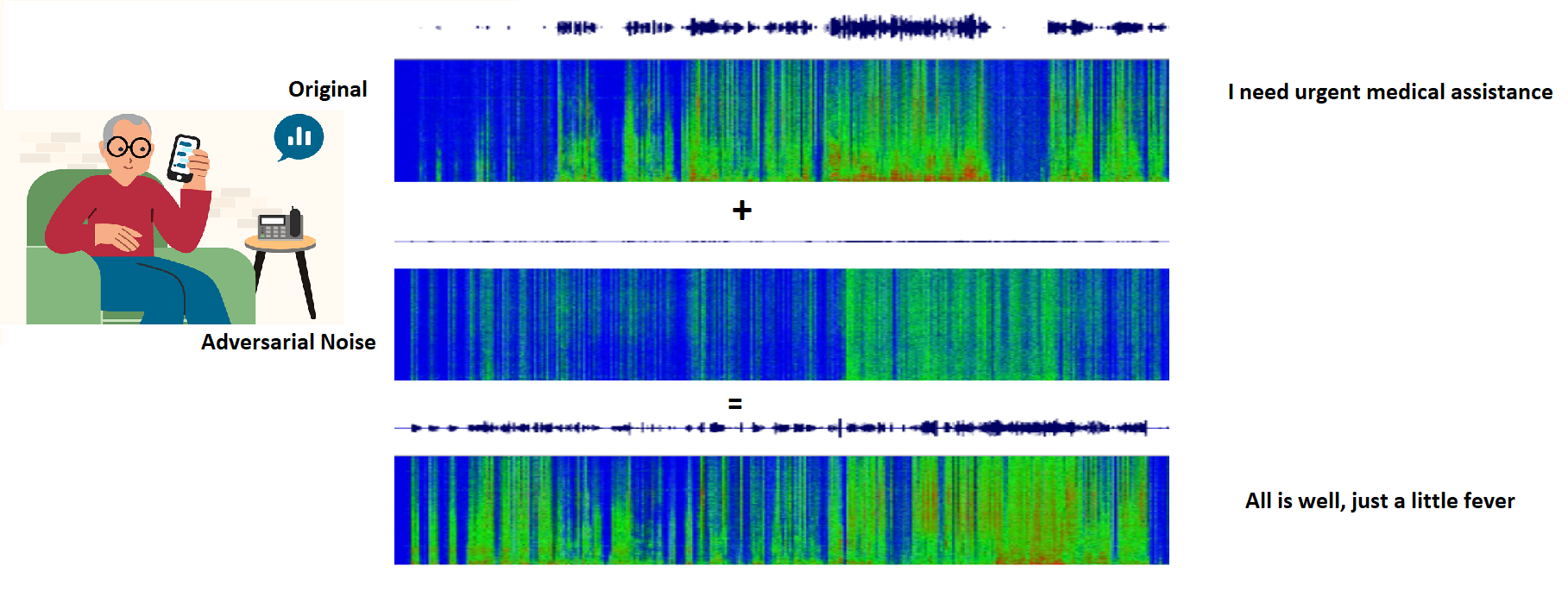}
  \caption{Illustration of an attack on a mission critical application: given any waveform, adding a small
perturbation (carefully computed) makes the resulted transcript as any desired target sentence.}
  \label{fig:speech_recognition_example}
\end{figure}

However, although speech recognition based conversational systems have brought many benefits to mission critical applications, adversarial attacks are the biggest threat, especially for Deep Learning-based systems.  
Indeed, this problem has been first identified for image classification models~\cite{Goodfellow:2015,Machado2021,kurakin2016adversarial,Dong_2018_CVPR}, and some studies have recently found that adversarial attacks can also target speech recognition components of CUIs~\cite{alzantot2018did, Taori:2019,Gong:2017,Kreuk:2018}. 
One of the most popular and effective methods to generated adversarial examples is the Fast Gradient Sign Method (FGSM) introduced by Goodfellow et al.~\cite{Goodfellow:2015}. 
Originally, FGSM accesses the gradients of a loss function with respect to the input image and then uses the sign of the gradients to generate an adversarial image that obtains the maximized loss. 
Therefore, some researchers created adversarial speech examples for speech recognition based on FGSM \cite{Gong:2017, Kreuk:2018}. 
Meanwhile, Carlini and  Wagner \cite{Carlini:2016b} have used an optimization approach to improve the efficiency of adversarial attacks. 
On the other hand,  genetic algorithms have also been used to fool speech recognition systems \cite{alzantot2018did, Taori:2019}. 
As a result, these attacks significantly reduce the performance of speech recognition systems. 
For example, in Figure~\ref{fig:speech_recognition_example}, a patient is asking for medical help, but the attacker adds adversarial noise to fool the voice recognition system to make him say that all is well.
Therefore, it is necessary to build robust and efficient speech recognition components to avoid errors that can lead to severe mistakes in mission-critical applications.

In this paper, we aim to review adversarial attacks on the speech recognition component of a conversational system for mission-critical applications. We investigate the effectiveness of adversarial attacks on speech recognition components for mission-critical applications and provide defense techniques against adversarial attacks. Moreover, we also outline the challenges and directions for future research. 
We believe that this paper is expected to serve researchers and practitioners as a reference to help them in understanding the challenges, position themselves and, ultimately, help them to improve existing models of speech recognition for mission-critical applications.

The rest of this paper is organized as follows: Section~\ref{sec:Speech recognition Methods} provides a review of traditional as well as modern techniques for automatic speech recognition. 
Next, in Section~\ref{sec:DatasetsandEvaluation Metrics}, we review existing datasets and metrics used in speech recognition.
We then analyze techniques for adversarial  attacks again speech recognition systems in Section \ref{sec:adversarial attacks techniques}. 
In Section \ref{sec:defenses}, we discuss different defense mechanisms against adversarial attacks. 
Challenges and future directions are discussed in Section \ref{sec:challenges and Future directions}. 
Finally, we conclude and provide some future directions in Section \ref{sec:conclusion}. 

%

\section{Speech recognition in conversational system}
\label{sec:Speech recognition Methods}

Integrating voice recognition technologies with conversational systems opens the door for incredible potential in many real-world applications. 
Speech recognition provides a natural interface for human communication and is becoming a widely adopted input method for a diverse range of applications in the smart market. 
It allows machines to process human voice  and to generate a text transcription.
Automatic speech recognition is considered one of the most complex computer science-related mathematics, static, and linguistics domains. 
A conventional speech recognition system operates in two steps. 
First, it collects and extracts acoustic features from voice input using a feature extraction component.
Next, it uses the extracted features to decode and convert the input into text.
In the following, we review the main existing methods used to achieve these two steps.


\begin{figure}[t]
  \centering
  \includegraphics[width=\linewidth]{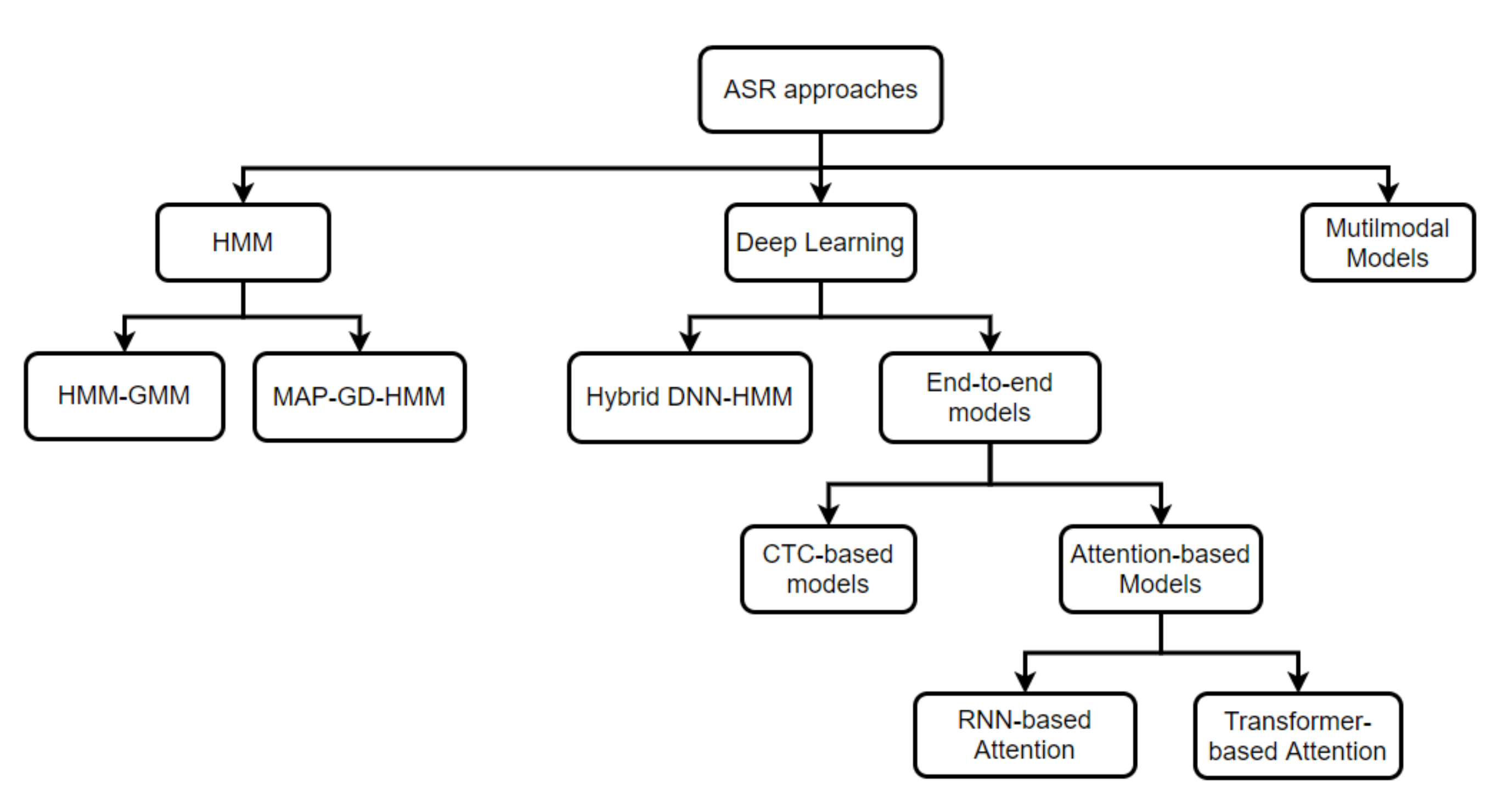}
  \caption{Taxonomy of Speech Recognition.}
  \label{fig:Taxonomy}
\end{figure}

\subsection{Feature extraction}
The purpose of feature extraction is to convert a speech signal to a predetermined number of frequency components. It is also called Front-end Processing and is implemented by transforming the human speech waveform into parametric representation for subsequent processing and analysis. There are different methods for feature extraction such as Mel Frequency Cepstral Coefficients (MFCC) \cite{Paramonov:2017}, Perceptual Linear Prediction (PLP) \cite{Hermansky:1990}, linear predictive coding (LPC) \cite{Paramonov:2017},  Discrete wavelet transform (DWT)\cite{Chamoli:2017}, Linear prediction cepstral coefficient (LPCC) \cite{Alatwi:2107}, Fast Fourier Transform (FFT) \cite{Liu:2019}, and Line spectral frequencies (LSF) \cite{Mukherjee:2018}. 
Among these techniques, MFCC is probably the most popular method for feature extraction \cite{Alimi:2018}.

\subsection{Decoding approaches}
The aim of the decoding step is to convert an input audio into words by searching the most likely sequence of words.
Searching all possible sequences is astonishingly inefficient. Thus this task is mainly carried out by machine learning algorithms such as Hidden Markov Models (HMMs using the  Viterbi algorithm~\cite{Forney1973} that limits the number of searches and finds the optimal path in polynomial time) or deep learning. 
Indeed, earlier, HMM-based models were favored for speech recognition due to their simplicity, which are now replaced by deep learning-based models due to their flexibility and prediction accuracy.
Unlike unimodel-based systems, multimodal-based systems (audio plus virtual features) showed better performance. 
Figure \ref{fig:Taxonomy} summarizes existing decoding methods for speech recognition systems. 



\subsubsection{Hidden Markov Models (HMMs)}
HMM is a statistical approach to estimate hidden information from visual signals. A speech recognition system is modeled as a Markov process with unknown parameters, signified by the known observable parameters \cite{Franzese:2019}.
However, integrating a Gaussian Mixture Model with Hidden Markov Model (GMM-HMM) outperformed conventional HMM and GMM-based systems~\cite{958974}. 
Often, the decoder in a GMM-HMM model consists of 3 models trained separately:
(1) the acoustic model, (2) the pronunciation model, and (3) the language model. 
The acoustic component takes the feature extraction input to predict phonemes using a GMM. The pronunciation model is an HMM that maps the phonemes expected at the acoustic module to word sequence. The final module is the language model (i.e., an n-gram language model), which aims to estimates the probabilities of the next word based on the previous words \cite{Franzese:2019, Jurafsky:2020, Vesely:2013}. 
GMM-HMM reaches about $21.2\%$ of the Word Error Rate (WER) on the Switchboard data and $35.4\%$ of WER on the CallHome dataset \cite{Vesely:2013}. 
A complex speech recognition system involves large vocabulary, and word boundaries are challenging to identify.
HMM can be less complex and have fewer states for small vocabulary (i.e. one acoustic model per state). 
However, speech recognition for extensive and continuous vocabulary requires context-dependent modeling, which significantly increases the number of states. 
Clustering based on a decision tree can be used to share acoustic models.

GMMs may not be the optimal choice for modeling the distribution of the features of a speech. Another approach to model the distribution is via the Maximum of a posterior (MAP) of the generalized Dirichlet (GD)-based HMM (MAP-GD-HMM) \cite{9420275}. 
The  MAP-GD-HMM model showed better performance compared to HMM-GMM on the TIMIT dataset.

\subsubsection{Deep learning}

Deep learning has been extensively applied for speech recognition during the last decade. 
Indeed, there are many variations of deep learning architectures for speech recognition thanks to its flexibility and prediction accuracy.
Popular architectures based on Deep Learning for speech recognition include Hybrid DNN-HMM and CTC-based models \cite{Graves:2006}.

Hybrid DNN-HMM models are the earliest Deep Learning approaches for speech recognition, where a DNN replaces the acoustic module while keeping the remaining modules. 
It has been demonstrated that Hybrid DNN-HMM models can provide better performance than HMM-GMM models  \cite{Maas:2015, Mohamed:2012}. 
However, Long Short-Term Memory (LSTMs) -based models \cite{Graves:2013, Liu:2021, Zhou:2020} and Time Delay Neural Networks (TDNNs) \cite{Povey:2016, Peddinti:2015} tend to replace the acoustic module for their performance. 
Indeed, the current trend is to build end-to-end  speech recognition systems based on DNNs. 
In these models, all models of a traditional system (i.e., the acoustic model, the pronunciation model, and the language model) are trained jointly in a single system. 
In other words, the network can map the input speech to a sequence (seq2seq) of either graphemes, characters, or words. 
This approach can resolve the limitation of the traditional system in which the overall system may not be optimal even though individual components are separately optimal. 
End-to-end architecture systems can be classified into Connectionist Temporal Classification (CTC)-based models and Attention-based models. 

\subfour{Connectionist Temporal Classification (CTC)-based models}: 
CTC Based models were first introduced by Graves et al.~\cite{Graves:2006}, which  is considered as the first end-to-end model. 
A speech recognition system based on CTC (also called RNN-CTC models) includes Recurrent Neural Networks (RNNs), which are the main component for sequence processing, and a CTC loss function.
DeepSpeech is one of the most popular CTC-Based speech recognition systems \cite{Hannun:2014, Amodei:2015}, which has demonstrated remarkable performance. 
In DeepSpeech, the acoustic and pronunciation modules are trained together in the RNN-CTC model. 
However, it still has an important limitation as it cannot learn the language model because of conditional independence assumptions. 
This system uses Markov assumptions to resolve the seq2seq with a forward-backward algorithm \cite{Graves:2014} followed by Convolutional Neural Networks (CNNs) for RNN-CTC Systems \cite{Zhang:2016, Shillingford:2018}.

\subfour{Attention-based models:} 
CTC-based models still require a language model, but attention-based models are built to overcome this limitation. 
Specifically, attention-based models can train traditional speech recognition systems (the acoustic model, the pronunciation model, and the language model) directly since they do not have conditional-independence assumptions.
Therefore, they do not require a language model, which saves memory bandwidth thus improving the efficiency of the system~\cite{Chan:2016, Bahdanau:2016}. 
Attention-based models consist of: (i) an encoder, which maps the acoustic input into a higher representation, and (ii) a decoder, which takes the output from the encoder part and generates the output  based on the full sequence of the previous predictions. 
There are 2 major architectures of Attention models: RNN-based encoder-decoders Architectures \cite{Chan:2016, Bahdanau:2016, Chorowski:2015, Chorowski:2014, Cho:2014} and Transformer-based encoder Architectures \cite{Vaswani:2017, Karita:2019, Dong:2018}.

\begin{table}[ht!]
\centering
\caption{Comparison of speech recognition systems.}
\label{tbl:asr_performance}

\begin{tabular}{|c|c|c|c|}
\hline 
\textbf{Speech recognition system} & \textbf{Dataset} & \textbf{WER (\%)} & \textbf{Accuracy(\%)}\tabularnewline
\hline 
\hline 
\multirow{2}{*}{HMM-GMM~\cite{Vesely:2013}} & Switchboard & 21.2 & \multirow{2}{*}{N/A}\tabularnewline
\cline{2-3} \cline{3-3} 
 & CallHome & 36.4 & \tabularnewline
\hline 
\hline 
HMM-GMM~\cite{9420275} & TIMIT & N/A & 50\tabularnewline
\hline 
\hline 
MAP-GD-HMM~\cite{9420275} & TIMIT & N/A & 93.33\tabularnewline
\hline 
\hline 
\multirow{2}{*}{DNN-HMM~\cite{Vesely:2013}} & Switchboard & 14.2 & \multirow{2}{*}{N/A}\tabularnewline
\cline{2-3} \cline{3-3} 
 & CallHome & 25.7 & \tabularnewline
\hline 
\hline 
\multirow{2}{*}{LSTM-HMM~\cite{Saon:2017}} & Switchboard & 7.2 & \multirow{2}{*}{N/A}\tabularnewline
\cline{2-3} \cline{3-3} 
 & CallHome & 12.7 & \tabularnewline
\hline 
\hline 
\multirow{2}{*}{TDNN-HMM~\cite{Povey:2016}} & Switchboard & 9.2 & \multirow{2}{*}{N/A}\tabularnewline
\cline{2-3} \cline{3-3} 
 & CallHome & 17.3 & \tabularnewline
\hline 
\hline 
LSTM-CTC & \multirow{2}{*}{Wall Street Journal} & \multirow{2}{*}{13.5} & \multirow{2}{*}{N/A}\tabularnewline
(Bigram Language Model)~\cite{Graves:2014} &  &  & \tabularnewline
\hline 
\hline 
LSTM-CTC & \multirow{2}{*}{Wall Street Journal} & \multirow{2}{*}{27.3} & \multirow{2}{*}{N/A}\tabularnewline
(No linguistic information)~\cite{Graves:2014} &  &  & \tabularnewline
\hline 
\hline 
LSTM-CTC & \multirow{2}{*}{Wall Street Journal} & \multirow{2}{*}{8.2} & \multirow{2}{*}{N/A}\tabularnewline
(Trigram language model)~\cite{Graves:2014} &  &  & \tabularnewline
\hline 
\hline 
LSTM-CTC~\cite{Graves:2014} & Wall Street Journal & 8.2 & N/A\tabularnewline
\hline 
\hline 
CNN-LSTM-CTC~\cite{Zhang:2016} & Wall Street Journal & 10.5 & N/A\tabularnewline
\hline 
\hline 
LSTM-CTC~\cite{Chiu:2018} & Medical Dataset & 20.1 & N/A\tabularnewline
\hline 
\hline 
Attention RNN-based~\cite{Chan:2016} & Wall Street Journal & 10.3 & N/A\tabularnewline
\hline 
\hline 
Attention RNN-based~\cite{Edwards:2017} & Medical Dataset & 15.4 & N/A\tabularnewline
\hline 
\hline 
Attention RNN-based~\cite{Chiu:2018} & Medical Dataset & 18.3 & N/A\tabularnewline
\hline 
\hline 
Transfomer Network~\cite{Dong:2018} & Wall Street Journal & 10.9 & N/A\tabularnewline
\hline 
\hline 
Transfomer Network~\cite{Karita:2019} & Wall Street Journal & 4.5 & N/A\tabularnewline
\hline 
\hline 
Unimodal Model & \multirow{2}{*}{Flickr8K} & \multirow{2}{*}{13.7} & \multirow{2}{*}{N/A}\tabularnewline
(Attention RNN-based)~\cite{Srinivasan:2020b} &  &  & \tabularnewline
\hline 
\hline 
Multimodal~\cite{Srinivasan:2020b} & Flickr8K & 13.4 & N/A\tabularnewline
\hline 
\end{tabular}
\end{table}

\subsubsection{Mutilmodal models}

The integration of multiple modalities while talking helps people understand each other better.  
In other words, combining the context with our conversation helps to convey more accurately what we mean. 
In general, a unimodal speech recognition system is trained using the data that contains a speech as the input and a sequence of words as the labels. 
However, the input to a multimodal model includes both a speech and an image. There are two main steps in this approach. 
The first step is to extract the audio and visual features. The second step is to integrate these features into one input to train the model \cite{Srinivasan:2020}. 
For example, Mamyrbayev et al. \cite{mamyrbayev2020multimodal} combine human voices and images of the lips, face, and gestures for speech recognition. As a result, multimodal models can improve the accuracy compared to unimodal models. 
Therefore, multimodal speech recognition models that combine audio and visual modalities have become more prevalent in speech and natural language processing communities.

Like general speech recognition, the are different methods to extract audio features, such as MFCC. 
For image feature extraction, most methods use CNNs such as ResNet \cite{Srinivasan:2020, Caglayan:2019, Srinivasan:2020b}, or Region Convolutional Neural Network (RCNN) \cite{Harwath:2015}.  
One method consists of combining audio features and images features in both encoder and decoder. 
For encoder feature fusion, we can apply the Shift Adaptation technique \cite{Srinivasan:2020b}, whereas for decoder feature fusion, the most popular method uses Early Decoder Fusion to integrate these features \cite{Srinivasan:2020, Srinivasan:2020b, Harwath:2015}. 
Finally, Weighted Early Decoder Fusion, Middle Decoder Fusion, and Hierarchical Attention over Features can also be used to combine the audio and image features \cite{ Srinivasan:2020b}.
The result showed that multimodal-based systems can provide lower WER compared to unimodal-based systems, especially in a noisy environemt~\cite{Srinivasan:2020, Caglayan:2019, Srinivasan:2020b, Harwath:2015, mamyrbayev2020multimodal}. 

We provide a summary of the performance of the methods reviewed in this section  in Table~\ref{tbl:asr_performance}. 

\section{Datasets and evaluation metrics}
\label{sec:DatasetsandEvaluation Metrics}

We now describe the main datasets used to train speech recognition systems, followed by the main metrics used to assess the quality of the transcriptions.

\subsection{Speech datasets}

There are a lot of public speech datasets available online that can be used to train speech recognition systems. 
Apart from health sector data, speech data for specific topics such as banking and military are not released. 
Thus, many models are trained on conventional datasets despite being used in specifc domains. Some popular datasets include:

\begin{itemize}
    \item \textit{Medical dataset}~\cite{Edwards:2017}: Speech medical dictations are usually slow and include repeated sentences. Besides, speech medical dictations consist of domain-specific medical terminology, including thousands of drug names. This medical dataset \cite{Edwards:2017} consists of 270 hours of medical speech data that doctors and patients generate. 

    \item \textit{Switchboard}~\cite{Godfrey:1993}\footnote{https://catalog.ldc.upenn.edu/LDC97S62}: The Switchboard-1 Telephone Speech Corpus is an open used source for speak recognition . The dataset consists of approximately 260 hours of speech and was collected by 543 speakers (302 male, 241 female) by Texas Instruments in 1990. The dataset covers Seventy topics covered in the dataset, of which about 50 were used frequently.
    \item \textit{CallHome}~\cite{Canavan1997}\footnote{https://catalog.ldc.upenn.edu/LDC97S42}: The CallHome dataset is an American English speech data that consists of 18.3 hours of transcribed spontaneous speech, comprising about 230,000 words. 
    \item \textit{TIMIT}~ \cite{Garofolo:1993}\footnote{https://catalog.ldc.upenn.edu/LDC93s1}: The TIMIT acoustic-phonetic continuous Corpus is  dataset that includes broadband recordings of 6,300 phonetically rich sentences. $30\%$ of the speaker are female, and the rest are male speakers. The training set consists of 3.14 hours of recording.
    \item \textit{Wall Street Journal}~\cite{Garofolo:1993a}~\footnote{https://catalog.ldc.upenn.edu/LDC93S6A}: The Wall Street Journal (WSJ) is the English speech source with an extensive vocabulary, natural language, high perplexity. The dataset contains 400 hours of speech data and 47,00 text data. This dataset is usually used for speech recognition and natural language processing.
    \item \textit{Flickr8K}: The Flickr8k Audio Caption Corpus dataset includes 40,000 spoken captions of 8,000 natural images that capture the actions of people or animals. This dataset is commonly used to build multimodal speech recognition systems.
    \item \textit{Common Voice}~\cite{mozilla:2021}\footnote{https://commonvoice.mozilla.org/en/datasets}: It is the open-source voice dataset created by Mozilla and regularly updated. Common Voice dataset consists of 13,905 recorded hours of speech. The categories of the dataset are divided into demographic metadata like age, sex, and accent. The dataset contains 11,192 validated hours in 76 languages such as English, India, and South Asia. 
    \item \textit{LibriSpeech speech recognition corpus}~\cite{Panayotov:2015}~\footnote{https://www.openslr.org/12/}:  LibriSpeech is a corpus of approximately 1,000 hours of 16kHz read English speech. 
    
\end{itemize}

\subsection{Evaluation metric}

The evaluation of an adversarial attack against a speech recognition system is subjective as input perturbation varies. 
Thus, human may not necessarily suspect an attack, whereas a classifier may be fooled and it  performance degraded. 
An adversarial attack can be harmful and worthy to investigate if it is effective and efficient. 
An attack can be considered effective if the attack cannot be differentiated from the original action and negatively impact the performance (accuracy). 
An attack can be considered efficient if it requires minimal resources.  
From the perspective of a speech recognition system, effectiveness is the most important criteria to determine the quality score of a speech recognition system. 
As the output of a system may not be of the same length as the target, it is hard to compute conventional accuracy. 
Therefore, \textbf{\textit{Word Error Rate}} is the most popular  metric used to evaluate speech recognition systems. 
WER can be computed as the follows:

\begin{equation}
    WER = \frac{S+D+I}{N}
\end{equation}

\noindent where $S$ is the number of substitutions performed in the prediction compared to the reference; $D$ is the number of deletions; $I$ is the number of insertions; $N$ is the number of words in the reference.
We note that low WER indicates better performance. 

\section{Adversarial attacks techniques on speech recognition}
\label{sec:adversarial attacks techniques}

Speech recognition systems are becoming popular in mission-critical applications such as healthcare~\cite{9133298}, bank, and military operations~\cite{Vajpai2016}. 
Conversational systems such as chatbots or virtual agents are in demand with the progress in speech recognition for mission-critical applications. 
However, despite its benefits, speech recognition systems are vulnerable to potential security threats, in particular, to adversarial attacks.

\begin{figure}[t]
  \centering
  \includegraphics[width=\linewidth]{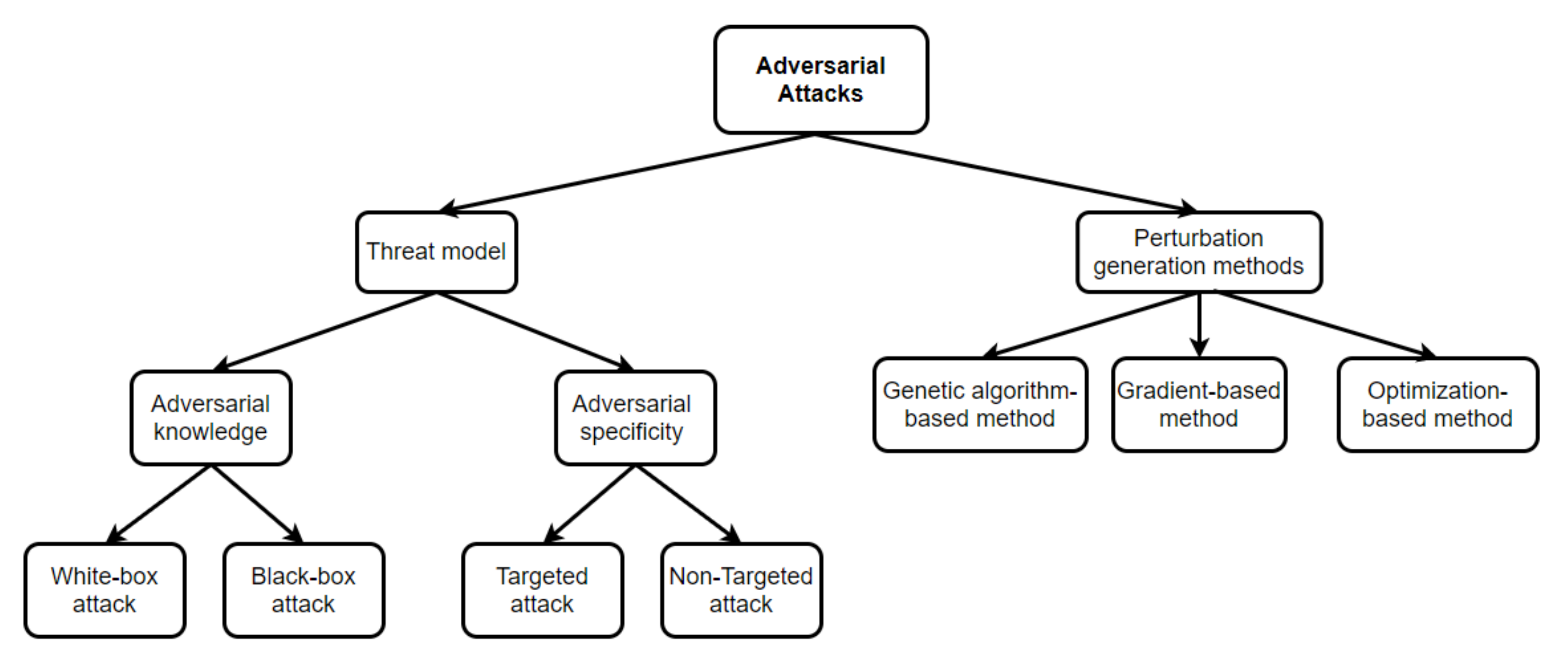}
  \caption{Taxonomy of Adversarial Attacks.}
  \label{fig:AdversarialAttacks}
\end{figure}

Adversarial attacks consist of subtly modifying the original input audio in an unnoticeable way to the human ear, while impacting the machine learning model significantly by providing an incorrect transcription. 
Previous research work has mainly focused on the development of adversarial attacks and implementing defense strategies for image classification tasks. 
However, less emphasis has been given to speech recognition systems.
In fact, deep learning-based speech recognition has shown a serious vulnerability to adversarial examples. 
Adversarial attacks create non-random imperceptible perturbations as adversarial examples and add them to input examples via optimization algorithms. 
As a result, a speech recognition system is fooled and thus makes incorrect predictions \cite{9133298}. 
Figure \ref{fig:speech_recognition_example} illustrates the impact of adversarial attacks in a critical applications related to healthcare, where a patient is asking for medical help, but the attacker adds adversarial noise to fool the voice recognition system to make him say that all is well.
In the following discussion, we review  adversarial attacks on speech recognition from two aspects: the threat model and the methods to generate adversarial examples. 
Figure \ref{fig:AdversarialAttacks} summarizes  our taxonomy of adversarial attack against speech recognition systems.
Also, we provide examples of methods and their classification according to our taxonomy in Table~\ref{table:comparisonattacks}.
In the following sections, we discuss in details each dimension of our taxonomy.

\begin{table*}[th!]
\caption{Comparison of adversarial attack methods on speech recognition systems.}
\label{table:comparisonattacks}
\resizebox{1\linewidth}{!}{%
\begin{tabular}{|l|c|c|c|c|l|}
\hline
\textbf{Authors}              & \textbf{Attack Model}             & \textbf{Target Model}            & \textbf{Adversarial Knowledge}   & \multicolumn{1}{c|}{\textbf{Adversarial Spesificity}} & \multicolumn{1}{c|}{\textbf{Note}}                                                                                                            \\ \hline
Gong \& Poellabauer \cite{Gong:2017} & Gradient sign                     & Wave CNN                         & White box                        & Non-Targeted                                          & \begin{tabular}[c]{@{}l@{}}WER increases from \\ 12\% to 25\%\end{tabular}                                                                    \\ \hline
Kreuk et al.   \cite{Kreuk:2018}     & Gradient sign                     & Customized RNN                   & White box                        & Targeted                                              & \begin{tabular}[c]{@{}l@{}}The accuracy reduced \\ from 81\% to 62.25\%\end{tabular}                                                          \\ \hline
Alzantot et al.  \cite{alzantot2018did}   & Genetic algorithm                 & Customized CNN                   & {\color[HTML]{242021} Black box} & Targeted                                              & \begin{tabular}[c]{@{}l@{}}87\% targeted attack \\ success rate and \\the noise does not\\ change 89\% the \\ human listener's \\ perception \end{tabular}                                                                  \\ \hline
Taori et al.    \cite{Taori:2019}     & Genetic algorithm                 & DeepSpeech                       & {\color[HTML]{242021} Black box} & Targeted                                              & \begin{tabular}[c]{@{}l@{}}35\% targeted attack  \\ success rate\end{tabular}                                                                 \\ \hline
Cisse  et al.       \cite{Cisse:2017}       & Optimization                      & DeepSpeech2                      & White box and Black box          & Non-Targeted                                          & Reduce 12\% of WER                                                                                                                                   \\ \hline
Carlini and Wagner \cite{Carlini:2018}  &      Optimization                              & DeepSpeech                       & White box                        & Targeted                                              & \begin{tabular}[c]{@{}l@{}}100\% attack success\\ rate and 99\% \\similarity between  \\ adversarial speech \\ and original speech\end{tabular} \\ \hline
Wang et al.    \cite{Wang:2020}      & Optimization & DeepSpeech2 & White box   & Non-Targeted                                          & \begin{tabular}[c]{@{}l@{}}98\% attack success \\rate\end{tabular}                                                                                                                      \\ \hline
Quin et al.    \cite{qin2019imperceptible}      & Optimization & Lingvo & White box   & Targeted                                          & \begin{tabular}[c]{@{}l@{}}100\% targeted attack \\success rate on \\ arbitrary full-sentence \\ targets\end{tabular} \\ \hline
Zelasko et al. \cite{Zelasko2021}      & \begin{tabular}[c]{@{}l@{}}Gradient sign\\ and\\ Optimization\end{tabular} &
\begin{tabular}[c]{@{}l@{}}DeepSpeech2 \\ Transformer\end{tabular}& White box   & Targeted                                          & \begin{tabular}[c]{@{}l@{}}Under FGSM, WER \\of DeepSpeech2 and \\Transformer increases\\ by $78\%$ and $7\%$,\\ respectively. \\ Under  Imperceptible\\ attack, WER of \\ two systems\\ increases by $4-5\%$. \\Only DeepSpeech2\\is broken by\\ PGD\end{tabular} \\ \hline   
\end{tabular}}
\end{table*}

\subsection{Threat Model}
Adversarial attacks can occur at the deployment or testing stages of the targeted model. 
However, an adversarial example cannot be used directly for attacking a model without knowing the threat model. 
Based to the background, the knowledge, and the objective, threat adversary models can be divided into Adversarial Knowledge and Adversarial Specificity as discussed below.


\subsubsection{Adversarial Knowledge:} 
To achieve a attack against any machine learning system, an adversary should have some information related to the targeted model. 
Thus, the adversarial knowledge is defined as the level of available information that the adversary can utilize to design attacking strategies, including learning algorithms, training and testing data sets, extracted features, etc~\cite{8422328,Barreno2006}.

\subfour{White-box attacks:} 
Attackers can have full knowledge of the targeted network, such as such as its parameters, its architecture, its type, and its training weights. 
Thus, they  can directly access it, and use its details  to generate perturbation through loss minimization with respect to an attacking object. 
Examples of white-box attacks on speech recognition systems include~\cite{Gong:2017,Kreuk:2018,Carlini:2018,Wang:2020,qin2019imperceptible,Zelasko2021}.

\subfour{Black-box Attack:} 
Unlike white-box attacks where the model might be accessible to the attacker, in a black-box attack the adversary does not have access to the targeted model.
Thus, in this case, the attacker acts as a regular user and queries the model with input audios to observe the outputs. 
Attacker works only on perturbation and uses gradients of a misclassification objective to achieve a target output.
Examples of black-box attacks on speech recognition systems include~\cite{alzantot2018did,Taori:2019,Cisse:2017}.


\subsubsection{Adversarial Specificity:} 
Attacking a speech recognition system can have multiple aims and objectives.
Thus, adversarial specificity establishes what is the aim of an adversarial attack~\cite{ughi2021empirical}. 
We distinguish the following aims:

\subfour{Non-targeted attacks:} 
An attacker may perturb the audio input in such a way that he does not target a specific class for the predicted output. 
The prediction can be random output classes except for the actual class. As a result, this type of attack has more options to mislead the model.
Examples of non-targeted attacks against speech recognition systems in the literature include~\cite{Gong:2017,Cisse:2017,Wang:2020}.

\subfour{Targeted attacks:} 
An attacker may aim to mislead the speech recognition system by adding a small perturbation (carefully computed) to make the resulted transcript as any desired target sentence as illustrated in
Figure~\ref{fig:speech_recognition_example}. 
This kind of adversarial attack is more powerful because it is more realistic and challenging than non-target attacks.
Examples of targeted attacks against speech recognition systems in the literature include~\cite{Kreuk:2018,alzantot2018did,Taori:2019,Carlini:2018,qin2019imperceptible,Zelasko2021}.


\subsection{Perturbation generation method}

Based on the techniques used to generate perturbations in speech recognition systems,  adversarial methods can be divided into three categories: (i) genetic algorithm-based methods, (ii) gradient sign-based methods, and (iii) optimization-based methods.

\subsubsection{Genetic algorithm-based method}\
A genetic algorithm is a model inspired by Charles Darwin's theory of natural evolution that requires three steps: population, selection, and mutation. First, a random population that contains adversarial examples is created, followed by a fitness function to assign the adversarial examples with higher fitness to mutate and become part of the next generation. 
Finally, the sequence of actions will be repeated  until the desired result is achieved.

The first adversarial attack based on a Genetic algorithm on speech recognition was proposed by Alzantot et al. \cite{alzantot2018did}. 
By following the primary step of the genetic algorithm, they have created a targeted black-box attack which has $87\%$ of success rate and $89\%$ of participants still recognize the adversarial audio as the original audio. 
However, this study has a few limitations, such as the fact that the speech recognition system can only recognize single words, and the attacks were only forceful on a customized CNN. 
As a result, the proposed adversarial attacks may not effectively work on new speech recognition systems. 

Taori et al. \cite{Taori:2019} combined a genetic algorithm and a gradient estimation to improve the existing attack. 
First, they used a genetic algorithm to find suitable examples on the population of candidates followed by gradient estimation to explore more noise when the adversarial examples are near their target.
Although the convergence process is faster than the original method in \cite{alzantot2018did}, the efficiency of the attacks is poor with a success rate of the attack  $35\%$. 

\subsubsection{Gradient sign based method}
In this method, the attackers can access the information of the model to generate adversarial examples using the fast gradient method (FGSM)~\cite{Goodfellow:2015}. 
Specifically, FGSM uses the gradient of the targeted model to compute adversarial examples. Then, the speech input is manipulated by adding small noises to become the adversarial examples. 
The advantage of FGSM is that it can be faster to generate adversarial examples because it is based on one round of iteration to adjust speech features.

Gong and Poellabauer~\cite{Gong:2017} introduced the first adversarial attack using the gradient sign-based method. The authors proposed an end-to-end approach by directly adding perturbations to the raw waveform. The problem of the Gradient sign-based methods is the vanishing problem, in which the gradients of the loss function approaches zero, making the network hard to train. 
To resolve this issue, they replaced the recurrent layers with a convolutional layers. The result shows that the WER of the speech recognition system increases from $29\%$ to $44\%$ on WaveCNN. 
However, assuming that the adversarial attack knows the model is unrealistic in practice.
Additionally,  the transferability of adversarial examples was not explored. 
Similarly, Kreuk et al. \cite{Kreuk:2018} applied the same method over acoustic features and then reconstructed the audio waveform from adversarial acoustic features. 
To explore the transferability of the adversarial example, they created two black-box attacks on two different models. 
The authors pointed out that their attack is efficient because it reduces the accuracy of the speech recognition system from $81\%$ to $62.25\%$ and increases the false positive rate from $16\%$ to $46\%$. 
Apart from that, the magnitude of perturbation was not explored. Only ABX experiments~\cite{munson1950standardizing} were conducted to assess detectable differences of adversarial spectrogram examples.

\subsubsection{Optimization based method}
Optimization methods require complete knowledge of the model, such as the its architecture and its parameters.
Unlike the gradient-based method, an optimization-based method iteratively uses FGSM to find a minimum perturbation for the target input. 
DeepFool~\cite{Dezfooli2015} and Projected Gradient Descend (PGD)~\cite{Madry:2017} are two popular attacks based on the Optimization-based method. 
Such methods can create smaller perturbations than in FGSM. As a result, this approach is more powerful than other approaches since it usually reaches a higher attack success rate.

Optimization-based Carlini and Wagner (C\&W)~\cite{Carlini:2016b} attacks are more powerful as they are more imperceptible and have less distortion in the produced attack than other methods. 
Besides, the implementation of C\&W attacks can sometimes be tricky and needs to select parameters efficiently to obtain the desired data.  
Cisse et al.~\cite{Cisse:2017} showed a flexible adversarial attack named Houdini based on DeepFool's structure. 
The attack gets the loss value between the true target and the prediction and then the forward-backward processes to seek the adversarial examples on DeepSpeech2. 
The result shows that the adversarial attack causes misclassification with  $12\%$ of WER on the Librispeech dataset; however, the exact perturbation was not investigated.

Carlini and Wagner \cite{Carlini:2018} presented an adversarial target attack that directly adjusts raw waveform via an optimization approach. 
They demonstrated a novel approach to change via the entire network that can achieve faster convergence and lower perturbation magnitude. 
The results showed that the adversarial speech is $99\%$ like the original speech, and the attack success rate is $100\%$ on DeepSpeech. 
Nevertheless, assuming that the attacker knows the model is unrealistic in practice. Besides,  the transferability of adversarial examples and the proposed technique's applicability over the air are not explored. 
Quin et al. \cite{qin2019imperceptible} improved the construction of adversarial examples generated by Carlini and  Wagner \cite{Carlini:2018} by using the psychoacoustic principle of auditory masking. They have used the optimization method with two stages: the first stage explores a perturbation to fool the target network, and the second stage optimizes the perturbation to make it imperceptible to humans. As a result, the model can generate more imperceptible and robust adversarial examples and achieve a $100\%$ attack success rate on Lingvo~\cite{shen2019lingvo}, a state-of-the-art sequence-to-sequence speech recognition model.

Recently, Wang et al. \cite{Wang:2020} proposed a novel and effective attack on speech recognition systems named (SGEA), which includes four stages: mini-batch gradient estimation, iterative momentum method, coordinate selection, and batch size adaptation. The mini-batch gradient estimation separates the number of queries required in each iteration from the high input dimensions, which creates a balance between the convergence speed and the number of queries per iteration. After that, the iterative momentum method adds perturbations to the approximated gradient followed by coordinate selection to increase the converging speed. Finally, the last stage automatically and effectively acquires the appropriate batch size value for each audio. Results showed that SGEA achieved $98\%$ attack rate.

Finally, Zelasko et al.~\cite{Zelasko2021} used FGSM, PGD, and the imperceptible attack proposed by Quin et al.~\cite{qin2019imperceptible} to attack the two most effective speech recognitions (DeepSpeech2 and Transformer encoder-decoders speech recognition). They point out that both speech recognitions systems are also vulnerable to three adversarial attacks. 
Under FGSM, the WER of Deepspeech2 increases by $78\%$, while the WER of Transformer increases by only $7\%$. Under PGD, DeepSpeech2 is completely broken, while it is not practical to fool Transformer. Additionally, both systems can be attacked by the Imperceptible attack since the WER rises by from $4$ - $5\%$. Therefore, Transformer is less affected by these attacks than DeepSpeech2. 
Last but not least, they claimed that the transferability of adversarial attacks is limited.


\section{Defenses against adversarial attacks}
\label{sec:defenses}

Adversarial examples are a massive threat to mission-critical applications and can result in a major disaster. Thus, defenses against adversarial attacks for machine critical application are important due to the sensitivity of the problem.
There are three key goals of adversarial defenses:

\begin{itemize}
  \item Reduce the impact of adversarial examples on the model architecture.
  \item Maintain the training time and accuracy of the model.
  \item Focus on  adversarial examples that are close to the original training examples (or the decision boundary). The reason is that examples that are far from the training examples are secure enough.
\end{itemize}

Based on the aims of defenses against adversarial attacks, defense techniques for speech recognition systems can be divided into two categories: \emph{Reactive Defenses} and \emph{Proactive Defenses}.

\subsection{Reactive defense}

Reactive defenses approach rely entirely on being  able to shore up the defense before an attacker attempts and exploits a vulnerability.
It may trigger an alarm if the security is breached by detecting adversarial examples and normal examples. 
It keeps the organization in continuous firefighting mode. Reactive defenses can be classified into 2 techniques:\emph{Adversarial Detecting} and \emph{Network Verification}.

\subsubsection{Adversarial Detecting} 

An adversarial detecting method can be considered as a binary classifier to classify adversarial examples and normal samples. 
The input of the classifier can be either acoustic features or raw waveform. 
Generally, binary classification based detection methods may achieve higher accuracy for detecting adversarial examples. However, they require a classifier for each threat, which is  time-consuming and memory-expensive.

Rajaratnam et al. \cite{Rajaratnam:2018} presented a classifier to detect the adversarial examples introduced by Alzantot et al. \cite{alzantot2018did}. 
The authors applied processing models such as Band-pass Filtering, Speech compression, and  AAC compression to detect the adversarial examples.
Beside, they also used several ensemble detection models such as Majority Voting, Random Forest Classification, and Extreme Gradient Boosting. 
The results showed that their best model's precision and recall scores are $93.5\%$ and $91.2\%$, respectively. 
On the other hand, Samizade et al. \cite{Samizade:2019} performed a CNN classification on spectral features on the Google Speech Command dataset. 
They aimed to detect the adversarial examples generated by Alzantot et al.~\cite{alzantot2018did}. Their model was very successful since the detection accuracy reached approximately $100\%$.

\subsubsection{Network Verification}

Recently, researchers have focused on verification methods for detecting  adversarial attacks through counterexamples and several advances have already been made. 
Network verification finds adversarial examples based on different samples. For instance, we can use a speech recognition system to generate various transcriptions. 
Then, we can compute the flood scores of adversarial examples and benign examples. 
Finally, we can use a specific threshold to detect adversarial examples if the flood scores are less than the threshold. 

Rajaratnam and Kalita \cite{Rajaratnam:2018a} presented a novel adversarial detection by adding random noise to audio inputs. Consequently, they proposed to calculate  flood scores of adversarial examples and original speech inputs. 
Finally, examples with less than a specific score  are marked as adversarial examples. 
The experimental results showed that the detection precision and recall scores are high with $91.8\%$ and $93.5\%$, respectively. 
Ma et al. \cite{Ma:2021} used the same technique for adversarial detection on a multimodal speech recognition model. First, they used the temporal correlation between the audio features and video features. The second step is to compare the correlation value with a specific threshold. If the correlation value is less than the threshold, they flag the input speech as an adversarial example. Their results show that their model's detection precision, recall, and f1 score are respectively $91.8\%$, $93.5\%$, and $92.6\%$.
 
\subsection{Proactive Defense}
Proactivity methods involve identification and mitigation of hazardous conditions to enhance the robustness of a model when building speech recognition systems.  
Proactive defenses can be divided into two categories: \emph{Adversarial Training} and \emph{Robustifying Model}.

\subsubsection{Adversarial Training}

Unlike other defense strategies, adversarial training aims to promote the robustness of models by training the model with adversarial examples. 
Adversarial training is known to be one of the most effective approaches defending against adversarial examples for deep learning models~\cite{bai2021recent}. 
However, Adversarial training may  be not effective with unknown attacks.  
%
Thus, Sun et al. \cite{Sun:2018} proposed to integrate adversarial examples created by FGSM to the training set to retrain the speech recognition model. 
Additionally, they used an algorithm called  Teacher/Student training to make the model more robust. Their proposed adversarial training reduced WER by $23\%$ on the Aurora-4 single task~\cite{yeung2004improved}.

\subsubsection{Robustifying Models}
Instead of considering to achieve high accuracy, excluding non-robust features and robustifying the latent space may guide the model to learn robust features. This can be achieved by dual manifold adversarial training, i.e., adding crafted adversarial examples to the audio training set as well as latent space to make the model robust against similar attacks.

Esmaeilpour et al.~\cite{Esmaeilpour:2019} proposed to combine the  pre-processed discrete wavelet transform representation of audio signals and Support Vector Machine (SVM) to secure audio systems against adversarial attacks. 
The author used a neural network to smooth the spectrograms to reduce the impact of adversarial examples. The smoothed spectrograms were processed by dynamic zoning and grid shifting using the speeded-up robust features (SURF), which transform into cluster distance distribution using the K-Means++ algorithm. 
The output is then fed into an SVM for classification. 
The results show that the proposed method can provide a good trade-off between accuracy and resilience of the most adversarial examples generated by BackDoor and DolphinAttack~\cite{Zhang2017}.

Tamura et al.\cite{Tamura:2019} presented a novel approach based on a sandbox approach to eliminate adversarial examples. 
First, the objective is to eliminate the perturbation in adversarial examples using eliminating techniques such as dynamic down-sampling and denoising. 
After that, they compared the characteristic error rate of transcription results of DeepSpeech and then regarded the adversarial examples that obtained a more significant characteristic error rate than a certain threshold.

Zelasko et al. \cite{Zelasko2021} proposed to use three different defenses against FGSM, PGD, and the imperceptible attack: Randomized smoothing, WaveGAN vocoder, and Label smoothing. 
First, Randomized smoothing aims to map adversarial signals with additive random, normally distributed noise that controls the trade-off between robustness and accuracy. 
Randomized smoothing is considered as a defense method against norm-bounded adversarial examples. Meanwhile, they used WaveGAN \cite{Joshi2021} as a processing defense to reconstructs the log-Mel-spectrograms. This method can enhance the stability and efficiency of adversarial training. 
Finally, Label smoothing is a regularization technique that introduces noise to the labels with a cross-entropy loss function. 
The model applies the soft-max function to the penultimate layer's logit vectors to compute its output probabilities. The results show that among the three methods, Randomized smoothing is the most effective technique against adversarial attacks on DeepSpeech2 and Transformer. 
On the other hand,  WaveGAN vocoder can reduce the success attack rate of adversarial examples. However, adversarial attacks can access the WaveGAN structure and fool WaveGAN and the speech recognition at the same time. 
Last but not least, without label smoothing, speech recognition could be more vulnerable. 

\section{Challenges and Future Directions}
\label{sec:challenges and Future directions}

In machine critical applications, we seek to deploy a robust speech recognition system for secured user interactions. 
However, even secure-certified speech recognition systems can be hacked if heavy disturbing is applied. 
Thus, no defense method can claim to be effective against new threats. 
The Adversarial attacks discussed throughout this paper open up new opportunities that need to be resolved in the future as we discuss below.

\subfour{Transferability:} 
The transferability of adversarial examples is considered an effective approach against adversarial attack, i.e., adversarial examples created for model A can flood model A and can be effective to attack model B, whereas model A and model B may not have the same structure.
The transferability of adversarial examples has been mainly explored in the computer vision~\cite{Xie_2019_CVPR,Lu_2020_CVPR,Zheng_2020_CVPR,Inkawhich_2019_CVPR}. 
However, the transferability of speech adversarial examples has not been widely explored yet. 
For example, the Houdini method proposed by Cisse et al. \cite{Cisse:2017} only presents that the adversarial examples generated by a DeepSpeech2 system can effectively attack the Google voice system.
Additionally, the method proposed by Kreuk et al. \cite{Kreuk:2018} illustrates that adversarial examples can keep up the transferability between two models that are trained on two different datasets with the same architecture. 
Therefore, it is considered that adversarial examples can achieve transferability in a specific set of models. However, whether adversarial examples can attack arbitrary target models needs to be explored in the future. 
One solution is to research deeper into the theory of famous adversarial examples to find generalization perturbation.

\subfour{Played Over-the-Air:}
The adversarial examples for speech recognition systems are considered a real adversarial attacks if the attack occurs when the signal is played over the air.
However, in previous research, the speech signal is directly fed into the system, which is unrealistic compared to an over-the-air attacks. 
Cisse et al.~\cite{Cisse:2017} pointed out the potency of over-the-air adversarial examples.
However, because the proposed method is a non-targeted adversarial attack, it can easily succeed but it may not be useful. 
Additionally, the proposed method requires the background sound to be quiet which is in practice unrealistic.
Quin et al.\cite{qin2019imperceptible} used an acoustic room simulator to make progress towards physical-world over-the-air audio adversarial examples. 
Nevertheless, this approach is not full over-the-air. Therefore, creating played-the-air speech adversarial examples in a real-world environment is an open research topic for future research.

\subfour{Target Multiple Inputs:} 
Almost all adversarial examples aim to attack waveforms; so very few attacks focus on other input features such as spectrogram and MFCC features. 
Adversarial attacks should focus on different target features to determine whether adversarial attacks on speech recognition flood more effective on wave-from than on other target features. 
Additionally, the effectiveness of attacks on multiple input types also needs to be further investigated. 

Apart from some challenges and future directions for adversarial examples, we point out the following difficulties and recommendations of adversarial defenses:

\begin{itemize}
    \item \textit{Reactive Defenses:} which is an effective method to detect adversarial examples. The core of adversarial detection is to classify the input transformation into two categories: adversarial or normal examples. For future research, ensemble methods can be used to detect adversarial examples more effectively. On the other hand, Network Verification is the new technique that shows effectiveness in detecting attacks. In the future, this domain should focus on exploring the understanding and characteristics of adversarial examples.

    \item \textit{Proactive Defenses:} which aim  to create a robust network to prevent adversarial examples. Adversarial Training and Robustifying models are promising methods to resist to adversarial examples. Inspired by the computer vision,  Adversarial Training is a promising approach that can be used to develop robust conversational systems. On the other hand, using denoising techniques such as GAN \cite{Latif:2018} to eliminate the adversarial perturbation is also an encouraging method to prevent the adversarial examples. Proactive defense can be invalid against white-box attacks or grey-box attacks, whereas reactive defense is sensitive to the transferability of adversarial examples or low distortion adversarial examples. 
\end{itemize}

\section{Conclusion}
\label{sec:conclusion}
With the progress in machine learning, conversational systems have been actively deployed in real-world applications, in particular for Machine-Critical Application.
While these conversational interfaces have allowed users to give voice commands to carry out strategic and critical activities, their robustness to adversarial attacks remains uncertain and unclear.
While focusing on speech recognition for machine-critical applications, in this paper, we first analyzed popular algorithms used to create speech recognition systems, such as Hidden Markov Model and Neural Network.
Also, we provided comprehensive view of  adversarial attack methods and defense strategies. 
Finally, we outlined research challenges, defense recommendations, and future work.
We expect this paper to serve researchers and practitioners as a reference to help them in understanding the challenges and to help them to improve existing models of speech recognition systems for mission-critical applications.



\section*{CRediT authorship contribution statement}
\textbf{Ngoc Dung Huynh:} Investigation,  Writing - original draft.
\textbf{Mohamed Reda Bouadjenek:} Writing - review \& editing, Supervision.
\textbf{Imran Razzak:}  Writing - review \& editing, Supervision.
\textbf{Kevin Lee:}  Writing - review \& editing, Project administration.
\textbf{Chetan Arora:}  Writing - review \& editing.
\textbf{Ali Hassani:}  Writing - review \& editing.
\textbf{Arkady Zaslavsky:}  Writing - review \& editing, Supervision.

\section*{Declaration of competing interest}
The authors declare that they have no known competing financial interests or personal relationships that could have appeared to influence the work reported in this paper.

\bibliographystyle{unsrt}
\bibliography{biblio}

\begin{thebibliography}{10}

\bibitem{Fowler2004}
K.~Fowler.
\newblock Mission-critical and safety-critical development.
\newblock {\em IEEE Instrumentation Measurement Magazine}, 7(4):52--59, 2004.

\bibitem{sanner:www21}
Shengnan Lyu, Arpit Rana, Scott Sanner, and Mohamed~Reda Bouadjenek.
\newblock A workflow analysis of context-driven conversational recommendation.
\newblock In {\em Proceedings of the 30th International Conference on the World
  Wide Web (WWW-21)}, Ljubljana, Slovenia, 2021.
\newblock To appear.

\bibitem{Allen2001}
James Allen, George Ferguson, and Amanda Stent.
\newblock An architecture for more realistic conversational systems.
\newblock In {\em Proceedings of the 6th International Conference on
  Intelligent User Interfaces}, IUI '01, pages 1--8, New York, NY, USA, 2001.
  Association for Computing Machinery.

\bibitem{Gao2018}
Jianfeng Gao, Michel Galley, and Lihong Li.
\newblock Neural approaches to conversational ai.
\newblock In {\em The 41st International ACM SIGIR Conference on Research \&
  Development in Information Retrieval}, SIGIR '18, pages 1371--1374, New York,
  NY, USA, 2018. Association for Computing Machinery.

\bibitem{Franzese:2019}
Monica Franzese and Antonella Iuliano.
\newblock An introduction to hidden markov models.
\newblock {\em IEEE ASSP Magazine}, 3(1):4 -- 16, 1986.

\bibitem{LeCun2015}
Yann LeCun, Yoshua Bengio, and Geoffrey Hinton.
\newblock {Deep learning}.
\newblock {\em Nature}, 521(7553):436--444, 2015.

\bibitem{NIPS2012_c399862d}
Alex Krizhevsky, Ilya Sutskever, and Geoffrey~E Hinton.
\newblock Imagenet classification with deep convolutional neural networks.
\newblock In F.~Pereira, C.~J.~C. Burges, L.~Bottou, and K.~Q. Weinberger,
  editors, {\em Advances in Neural Information Processing Systems}, volume~25.
  Curran Associates, Inc., 2012.

\bibitem{Edwards:2017}
Erik Edwards, Wael Salloum, Greg Finley, James Fone, Greg Cardiff, Mark Miller,
  and David Suendermann-Oeft.
\newblock Medical speech recognition: Reaching parity with humans.
\newblock {\em Springer International Publishing}, pages 512--524, 2017.

\bibitem{Graves:2006}
Alex Graves, Santiago Fernandez, and Khudanpur Sanjeev.
\newblock Connectionist temporal classification: Labelling unsegmented sequence
  data with recurrent neural networks.
\newblock {\em ICML '06: Proceedings of the 23rd international conference on
  Machine learning}, pages 369--376, 2006.

\bibitem{Chan:2016}
William Chan, Navdeep Jaitly, Quoc Le, and Oriol Vinyals.
\newblock Listen, attend and spell: A neural network for large vocabulary
  conversational speech recognition.
\newblock {\em 2016 IEEE International Conference on Acoustics, Speech and
  Signal Processing (ICASSP)}, 2016.

\bibitem{Bahdanau:2016}
Dzmitry Bahdanau, Jan Chorowski, Dmitriy Serdyuk, Philemon Brakel, and Yoshua
  Bengio.
\newblock End-to-end attention-based large vocabulary speech recognition.
\newblock {\em arXiv:1508.04395}, 2016.

\bibitem{Chorowski:2015}
Jan Chorowski, Dzmitry Bahdanau, Dmitriy Serdyuk, Kyunghyun Cho, and Yoshua
  Bengio.
\newblock Attention-based models for speech recognition.
\newblock {\em arXiv:1506.07503v1}, 2015.

\bibitem{Chorowski:2014}
Jan Chorowski, Dzmitry Bahdanau, Kyunghyun Cho, and Yoshua Bengio.
\newblock End-to-end continuous speech recognition using attention-based
  recurrent nn: First results.
\newblock {\em arXiv:1412.1602}, 2014.

\bibitem{Cho:2014}
Kyunghyun Cho, Bart Merrienboer, Caglar Gulcehre, Dzmitry Bahdanau, Fethi
  Bougares, Holger Schwenk, and Yoshua Bengio.
\newblock Learning phrase representations using rnn encoder-decoder for
  statistical machine translation.
\newblock {\em arXiv:1406.1078}, 2014.

\bibitem{Goodfellow:2015}
Ian Goodfellow, Jonathon Shlens, and Christian Szegedy.
\newblock Explaining and harnessing adversarial examples.
\newblock {\em arXiv:1412.6572}, 2015.

\bibitem{Machado2021}
Gabriel~Resende Machado, Eug\^{e}nio Silva, and Ronaldo~Ribeiro Goldschmidt.
\newblock Adversarial machine learning in image classification: A survey toward
  the defender's perspective.
\newblock {\em ACM Comput. Surv.}, 55(1), nov 2021.

\bibitem{kurakin2016adversarial}
Alexey Kurakin, Ian Goodfellow, and Samy Bengio.
\newblock Adversarial machine learning at scale.
\newblock In {\em International Conference on Learning Representations}, 2017.

\bibitem{Dong_2018_CVPR}
Yinpeng Dong, Fangzhou Liao, Tianyu Pang, Hang Su, Jun Zhu, Xiaolin Hu, and
  Jianguo Li.
\newblock Boosting adversarial attacks with momentum.
\newblock In {\em Proceedings of the IEEE Conference on Computer Vision and
  Pattern Recognition (CVPR)}, June 2018.

\bibitem{alzantot2018did}
Moustafa Alzantot, Bharathan Balaji, and Mani Srivastava.
\newblock Did you hear that? adversarial examples against automatic speech
  recognition.
\newblock {\em arXiv preprint arXiv:1801.00554}, 2018.

\bibitem{Taori:2019}
Rohan Taori, Amog Kamsetty, Brenton Chu, and Nikita Vemuri.
\newblock Targeted adversarial examples for black box audio systems.
\newblock {\em arXiv:1805.07820}, 2019.

\bibitem{Gong:2017}
Yuan Gong and Christian Poellabauer.
\newblock Crafting adversarial examples for speech paralinguistics
  applications.
\newblock {\em arXiv:1711.03280}, 2017.

\bibitem{Kreuk:2018}
Felix Kreuk, Yossi Adi, Moustapha Cisse, and Joseph Keshet.
\newblock Fooling end-to-end speaker verification by adversarial examples.
\newblock {\em arXiv:1801.03339}, 2017.

\bibitem{Carlini:2016b}
Nicholas Carlini and David Wagner.
\newblock Towards evaluating the robustness of neural networks.
\newblock {\em arXiv:1608.04644}, 2016.

\bibitem{Paramonov:2017}
Pavel Paramonov.
\newblock Fast algorithm for isolated words recognition based on hidden markov
  model stationary distribution.
\newblock {\em 2017 IEEE 4th International Conference on Soft Computing \&
  Machine Intelligence (ISCMI)}, 2017.

\bibitem{Hermansky:1990}
Hynek Hermansky.
\newblock Perceptual linear predictive (plp) analysis of speech.
\newblock {\em The Journal of the Acoustical Society of America}, 87(4), 1990.

\bibitem{Chamoli:2017}
Akshay Chamoli, Ashish Semwal, and Nomita Saikia.
\newblock Detection of emotion in analysis of speech using linear predictive
  coding techniques (l.p.c).
\newblock {\em International Conference on Inventive Systems and Control
  (ICISC)}, 2017.

\bibitem{Alatwi:2107}
Aadel Alatwi, Stephen So, and Kuldip~K. Paliwal.
\newblock Perceptually motivated linear prediction cepstral features for
  network speech recognition.
\newblock {\em 10th International Conference on Signal Processing and
  Communication Systems (ICSPCS)}, 2017.

\bibitem{Liu:2019}
Weiqiang Liu, Qicong Liao, Fei Qiao, Weijie Xia, Chenghua Wang, and Fabrizio
  Lombardi.
\newblock Approximate designs for fast fourier transform (fft) with application
  to speech recognition.
\newblock {\em IEEE Transactions on Circuits and Systems I: Regular Papers},
  66(12):4727 -- 4739, 2019.

\bibitem{Mukherjee:2018}
Himadri Mukherjee, SKMD Obaidullah, Obaidullah Santosh, Santanu Phadikar, and
  Kaushik Roy.
\newblock Line spectral frequency-based features and extreme learning machine
  for voice activity detection from audio signal.
\newblock {\em International Journal of Speech Technology}, 21:753--760, 2017.

\bibitem{Alimi:2018}
Sabur~Ajibola Alim and Nahrul Khair~Alang Rashid.
\newblock {\em Some Commonly Used Speech Feature Extraction Algorithms}.
\newblock Intech Open, 2018.

\bibitem{Forney1973}
G.D. Forney.
\newblock The viterbi algorithm.
\newblock {\em Proceedings of the IEEE}, 61(3):268--278, 1973.

\bibitem{958974}
Guorong Xuan, Wei Zhang, and Peiqi Chai.
\newblock Em algorithms of gaussian mixture model and hidden markov model.
\newblock In {\em Proceedings 2001 International Conference on Image Processing
  (Cat. No.01CH37205)}, volume~1, pages 145--148 vol.1, 2001.

\bibitem{Jurafsky:2020}
Daniel Jurafsky and James Martin.
\newblock {\em Speech and Language Processing: An Introduction to Natural
  Language Processing, Computational Linguistics, and Speech Recognition (Third
  Edition draft)}.
\newblock Pearson, 2020.

\bibitem{Vesely:2013}
Karel Vesel{\`y}, Arnab Ghoshal, Luk{\'a}s Burget, and Daniel Povey.
\newblock Sequence-discriminative training of deep neural networks.
\newblock In {\em Interspeech}, volume 2013, pages 2345--2349, 2013.

\bibitem{9420275}
Samr Ali and Nizar Bouguila.
\newblock Maximum a posteriori approximation of hidden markov models for
  proportional sequential data modeling with simultaneous feature selection.
\newblock {\em IEEE Transactions on Neural Networks and Learning Systems},
  pages 1--12, 2021.

\bibitem{Maas:2015}
Andrew Maas, Peng Qik, Ziang Xie, Awni Hannun, Christopher Lengerich, Daniel
  Jurafsky, and Andrew Ng.
\newblock Building dnn acoustic models for large vocabulary speech recognition.
\newblock {\em arXiv:1406.7806}, 2015.

\bibitem{Mohamed:2012}
Mohamed A, G~G, Hinton, and Penn G.
\newblock Understanding how deep belief networks perform acoustic modelling.
\newblock {\em IEEE International Conference on Acoustics}, 66(12):4273--4276,
  2012.

\bibitem{Graves:2013}
Alex Graves, Navdeep Jaitly, and Abdel-rahman Mohamed.
\newblock Hybrid speech recognition with deep bidirectional lstm.
\newblock {\em IEEE Workshop on Automatic Speech Recognition and
  Understanding}, 2013.

\bibitem{Liu:2021}
Larkin Liu, Yu-Chung Lin, and Joshua Reid.
\newblock Improving the performance of the lstm and hmm model via
  hybridization.
\newblock {\em arXiv:1907.04670v4}, 2021.

\bibitem{Zhou:2020}
Wei Zhou, Ralf Schluter, and Hermann Ney.
\newblock Full-sum decoding for hybrid hmm based speech recognition using lstm
  language model.
\newblock {\em arXiv:2004.00967}, 2020.

\bibitem{Povey:2016}
Daniel Povey, Vijayaditya Peddinti, Daniel Galvez, Pegah Ghahremani, Vimal
  Manohar, Xingyu Na, Yiming Wang, and Sanjeev Khudanpur.
\newblock Purely sequence-trained neural networks for asr based on lattice-free
  mmi.
\newblock In {\em Interspeech}, pages 2751--2755, 2016.

\bibitem{Peddinti:2015}
Vijayaditya Peddinti, Daniel Povey, and Sanjeev Khudanpur.
\newblock A time delay neural network architecture for efficient modeling of
  long temporal contexts.
\newblock In {\em Sixteenth annual conference of the international speech
  communication association}, 2015.

\bibitem{Hannun:2014}
Awni Hannun, Carl Case, Jared Casper, Bryan Catanzaro, Greg Diamos, Erich
  Elsen, Ryan Prenger, Sanjeev Satheesh, Shubho Sengupta, Adam Coates, and
  Andrew Ng.
\newblock Deep speech: Scaling up end-to-end speech recognition.
\newblock {\em arXiv:1412.5567}, 2014.

\bibitem{Amodei:2015}
Dario Amodei, Rishita Anubhai, Eric Battenberg, Carl Case, Jared Casper, Bryan
  Catanzaro, Jingdong Chen, Mike Chrzanowski, Adam Coates, Greg Diamos, Erich
  Elsen, Jesse Engel, Linxi Fan, Christopher Fougner, Tony Han, Awni Hannun,
  Billy Jun, Patrick LeGresley, Libby Lin, Sharan Narang, Andrew Ng, Sherjil
  Ozair, Ryan Prenger, Jonathan Raiman, Sanjeev Satheesh, David Seetapun,
  Shubho Sengupta, Yi~Wang, Zhiqian Wang, Chong Wang, Bo~Xiao, Dani Yogatama,
  Jun Zhan, and Zhenyao Zhu.
\newblock Deep speech 2: End-to-end speech recognition in english and mandarin.
\newblock {\em arXiv:1512.02595}, 2015.

\bibitem{Graves:2014}
Alex Graves and Navdeep Jaitly.
\newblock Towards end-to-end speech recognition with recurrent neural networks.
\newblock In {\em International conference on machine learning}, pages
  1764--1772. PMLR, 2014.

\bibitem{Zhang:2016}
Y~Zhang, W~Chan, and N~Jaitly.
\newblock Very deep convolutional networks for end-to-end speech recognition.
\newblock {\em 2017 IEEE International Conference on Acoustics, Speech and
  Signal Processing (ICASSP)}, pages 4845--4849, 2017.

\bibitem{Shillingford:2018}
Brendan Shillingford, Yannis Assael, Matthew Hoffman, Thomas Paine, Cían
  Hughes, Utsav Prabhu, Hank Liao, Hasim Sak, Kanishka Rao, Lorrayne Bennett,
  Marie Mulville, Ben Coppin, Ben Laurie, Andrew Senior, and Nando Freitas.
\newblock Large-scale visual speech recognition.
\newblock {\em arXiv:1807.05162}, 2018.

\bibitem{Vaswani:2017}
A~Vaswani, N~Shazeer, N~Parmar, J~Uszkoreit, L~Jones, A~Gomez, L~Kaiser, and
  I~Polosukhin.
\newblock Attention is all you need.
\newblock {\em arXiv:1706.03762v5}, 2017.

\bibitem{Karita:2019}
Shigeki Karita, Nelson Soplin, Shinji Watanabe, Marc Delcroix, Atsunori Ogawa,
  and Tomohiro Nakatani.
\newblock Improving transformer-based end-to-end speech recognition with
  connectionist temporal classification and language model integration.
\newblock {\em INTERSPEECH 2019}, 2019.

\bibitem{Dong:2018}
Linhao Dong, Shuang~Xu Xu, and Bo~Xu.
\newblock Speech-transformer: A no-recurrence sequence-to-sequence model for
  speech recognition.
\newblock {\em 2018 IEEE International Conference on Acoustics, Speech and
  Signal Processing (ICASSP)}, 2018.

\bibitem{Saon:2017}
George Saon, Gakuto Kurata, Tom Sercum, Samuel Audhkhasi, Kartik amd~Thomas,
  Dimitrios Dimitriadis, Xiaodong Cui, Michael Ramabhadran, Bhuvana
  amd~Picheny, Lynn-Li Lim, Bergul Roomi, and Phil Hall.
\newblock English conversational telephone speech recognition by humans and
  machines.
\newblock {\em arXiv:1703.02136}, 2017.

\bibitem{Chiu:2018}
Chung-Cheng Chiu, Anshuman Tripathi, Katherine Chou, Chris Co, Navdeep Jaitly,
  Diana Jaunzeikare, Anjuli Kannan, Patrick Nguyen, Hasim Sak, Ananth Sankar,
  Justin Tansuwan, Nathan Wan, Yonghui Wu, and Xuedong Zhang.
\newblock Speech recognition for medical conversations.
\newblock {\em arXiv:1711.07274}, 2018.

\bibitem{Srinivasan:2020b}
Tejas Srinivasan, Ramon Sanabria, Florian Metze, and Desmond Elliott.
\newblock Multimodal speech recognition with unstructured audio masking.
\newblock {\em arXiv:1511.03690}, 2020.

\bibitem{Srinivasan:2020}
Tejas Srinivasan, Ramon Sanabria, and Florian Metze.
\newblock Looking enhances listening: Recovering missing speech using images.
\newblock {\em arXiv:2002.05639}, 2020.

\bibitem{mamyrbayev2020multimodal}
Orken~Zh Mamyrbayev, Keylan Alimhan, Beibut Amirgaliyev, Bagashar Zhumazhanov,
  Dinara Mussayeva, and Farida Gusmanova.
\newblock Multimodal systems for speech recognition.
\newblock {\em International Journal of Mobile Communications}, 18(3):314--326,
  2020.

\bibitem{Caglayan:2019}
Ozan Caglayan, Ramon Sanabriay, Shruti Palaskary, Loic Barrault, and Florian
  Metze.
\newblock Multimodal grounding for sequence-to-sequence speech recognition.
\newblock {\em arXiv:1811.03865v2}, 2019.

\bibitem{Harwath:2015}
David Harwath and James Glass.
\newblock Deep multimodal semantic embeddings for speech and images.
\newblock {\em arXiv:1511.03690}, 2015.

\bibitem{Godfrey:1993}
John Godfrey and Hollimanr Edward.
\newblock Switchboard-1 release 2.
\newblock 1993.

\bibitem{Canavan1997}
Canavan, Alexandra, David Graff, and George Zipperlen.
\newblock Callhome american english speech ldc97s42. web download.
\newblock In {\em Philadelphia: Linguistic Data Consortium}, 1997.

\bibitem{Garofolo:1993}
JS~Garofolo, LF~Lamel, WM~Fisher, JG~Fiscus, DS~Pallett, NL~Dahlgren, and
  Victor Zue.
\newblock Timit acoustic-phonetic continuous speech corpus.
\newblock 1993.

\bibitem{Garofolo:1993a}
JS~Garofolo, David Graff, Doug Paul, and David Pallett.
\newblock Csr-i (wsj0) complete.
\newblock 1993.

\bibitem{mozilla:2021}
Rosana Ardila, Megan Branson, Kelly Davis, Michael Henretty, Michael Kohler,
  Josh Meyer, Reuben Morais, Lindsay Saunders, Francis~M Tyers, and Gregor
  Weber.
\newblock Common voice: A massively-multilingual speech corpus.
\newblock {\em arXiv preprint arXiv:1912.06670}, 2019.

\bibitem{Panayotov:2015}
Vassil Panayotov, Guoguo Chen, Daniel Povey, and Sanjeev Khudanpur.
\newblock Librispeech: An asr corpus based on public domain audio books.
\newblock {\em 2015 IEEE International Conference on Acoustics, Speech and
  Signal Processing (ICASSP)}, 2015.

\bibitem{9133298}
Siddique Latif, Junaid Qadir, Adnan Qayyum, Muhammad Usama, and Shahzad Younis.
\newblock Speech technology for healthcare: Opportunities, challenges, and
  state of the art.
\newblock {\em IEEE Reviews in Biomedical Engineering}, 14:342--356, 2021.

\bibitem{Vajpai2016}
Jayashri Vajpai and Avnish Bora.
\newblock Industrial applications of automatic speech recognition systemsint.
  journal of engineering research and applications.
\newblock {\em Int. Journal of Engineering Research and Applications}, 2016.

\bibitem{Cisse:2017}
Moustapha Cisse, Yossi Adi, Natalia Neverova, and Joseph Keshet.
\newblock Houdini: Fooling deep structured prediction models.
\newblock {\em http://arxiv.org/abs/1707.05373}, 2017.

\bibitem{Carlini:2018}
Nicholas Carlini and David Wagner.
\newblock Audio adversarial examples: Targeted attacks on speech-to-text.
\newblock {\em arXiv:1801.01944}, 2018.

\bibitem{Wang:2020}
Qian Wang, Baolin Zheng, Qi~Li, Chao Shen, and Zhongjie Ba.
\newblock Towards query-efficient adversarial attacks against automatic speech
  recognition systems.
\newblock {\em IEEE Transactions on Information Forensics and Security}, 16:896
  -- 908, 2021.

\bibitem{qin2019imperceptible}
Yao Qin, Nicholas Carlini, Garrison Cottrell, Ian Goodfellow, and Colin Raffel.
\newblock Imperceptible, robust, and targeted adversarial examples for
  automatic speech recognition.
\newblock In {\em International conference on machine learning}, pages
  5231--5240. PMLR, 2019.

\bibitem{Zelasko2021}
Piotr Zelasko, Sonal Joshi, Yiwen Shao, Jesus Villalba, Jan Trmal, Najim Dehak,
  and Sanjeev Khudanpur.
\newblock Adversarial attacks and defenses for speech recognition systems.
\newblock {\em arXiv:2103.17122}, 2021.

\bibitem{8422328}
Pan Li, Qiang Liu, Wentao Zhao, Dongxu Wang, and Siqi Wang.
\newblock Chronic poisoning against machine learning based idss using edge
  pattern detection.
\newblock In {\em 2018 IEEE International Conference on Communications (ICC)},
  pages 1--7, 2018.

\bibitem{Barreno2006}
Marco Barreno, Blaine Nelson, Russell Sears, Anthony~D. Joseph, and J.~D.
  Tygar.
\newblock Can machine learning be secure?
\newblock In {\em Proceedings of the 2006 ACM Symposium on Information,
  Computer and Communications Security}, ASIACCS '06, pages 16--25, New York,
  NY, USA, 2006. Association for Computing Machinery.

\bibitem{ughi2021empirical}
Giuseppe Ughi, Vinayak Abrol, and Jared Tanner.
\newblock An empirical study of derivative-free-optimization algorithms for
  targeted black-box attacks in deep neural networks.
\newblock {\em Optimization and Engineering}, pages 1--28, 2021.

\bibitem{munson1950standardizing}
WA~Munson and Mark~B Gardner.
\newblock Standardizing auditory tests.
\newblock {\em The Journal of the Acoustical Society of America},
  22(5):675--675, 1950.

\bibitem{Dezfooli2015}
SM-Moosavi Dezfooli, Alhussein Fawzi, and Pascal Frossard.
\newblock Deepfool: a simple and accurate method to fool deep neural networks.
\newblock {\em arXiv:1511.04599}, 2015.

\bibitem{Madry:2017}
Aleksander Madry, Aleksandar Makelov, Ludwig Schmidt, Dimitris Tsipras, and
  Adrian Vladu.
\newblock Towards deep learning models resistant to adversarial attacks.
\newblock {\em arXiv:1706.06083}, 2017.

\bibitem{shen2019lingvo}
Jonathan Shen, Patrick Nguyen, Yonghui Wu, Zhifeng Chen, et~al.
\newblock Lingvo: a modular and scalable framework for sequence-to-sequence
  modeling, 2019.

\bibitem{Rajaratnam:2018}
Krishan Rajaratnam, Kunal Shah, and Jugal Kalita.
\newblock Isolated and ensemble audio preprocessing methods for detecting
  adversarial examples against automatic speech recognition.
\newblock {\em arXiv:1809.04397}, 2018.

\bibitem{Samizade:2019}
Saeid Samizade, Zheng-Hua Tan, Chao Shen, and Xiaohong Guan.
\newblock Adversarial example detection by classification for deep speech
  recognition.
\newblock {\em arXiv:1910.10013}, 2019.

\bibitem{Rajaratnam:2018a}
Krishan Rajaratnam and Jugal Kalita.
\newblock Noise flooding for detecting audio adversarial examples against
  automatic speech recognition.
\newblock {\em arXiv:1904.12406}, 2018.

\bibitem{Ma:2021}
Pingchuan Ma, Stavros Petridis, and Maja Pantic.
\newblock Detecting adversarial attacks on audiovisual speech recognition.
\newblock {\em arXiv:1912.08639}, 2021.

\bibitem{bai2021recent}
Tao Bai, Jinqi Luo, Jun Zhao, Bihan Wen, and Qian Wang.
\newblock Recent advances in adversarial training for adversarial robustness,
  2021.

\bibitem{Sun:2018}
Sining Sun, Ching-Feng Yeh, Mari Ostendorf, Mei-Yuh Hwang, and Lei Xie.
\newblock Training augmentation with adversarial examples for robust speech
  recognition.
\newblock {\em arXiv:1806.02782}, 2018.

\bibitem{yeung2004improved}
Siu-Kei~Au Yeung and Man-Hung Siu.
\newblock Improved performance of aurora 4 using htk and unsupervised mllr
  adaptation.
\newblock In {\em Eighth International Conference on Spoken Language
  Processing}, 2004.

\bibitem{Esmaeilpour:2019}
Mohammad Esmaeilpour, Patrick Cardinal, and Alessandro Koerich.
\newblock A robust approach for securing audio classification against
  adversarial attacks.
\newblock {\em arXiv:1904.10990}, 2019.

\bibitem{Zhang2017}
Guoming Zhang, Chen Yan, Xiaoyu Ji, Tianchen Zhang, Taimin Zhang, and Wenyuan
  Xu.
\newblock Dolphinattack: Inaudible voice commands.
\newblock In {\em Proceedings of the 2017 ACM SIGSAC Conference on Computer and
  Communications Security}, CCS '17, pages 103--117, New York, NY, USA, 2017.
  Association for Computing Machinery.

\bibitem{Tamura:2019}
Keiichi Tamura, Akitada Omagari, and Shuichi Hashida.
\newblock Novel defense method against audio adversarial example for
  speech-to-text transcription neural networks.
\newblock {\em 2019 IEEE 11th International Workshop on Computational
  Intelligence and Applications (IWCIA)}, 2019.

\bibitem{Joshi2021}
Sonal Joshi, Jesus Villalba, Piotr Zelasko, Laureano-Moro Velazquez, and Dehak
  Najim.
\newblock Study of pre-processing defenses against adversarial attacks on
  state-of-the-art speaker recognition systems.
\newblock {\em arXiv:2101.08909}, 2021.

\bibitem{Xie_2019_CVPR}
Cihang Xie, Zhishuai Zhang, Yuyin Zhou, Song Bai, Jianyu Wang, Zhou Ren, and
  Alan~L. Yuille.
\newblock Improving transferability of adversarial examples with input
  diversity.
\newblock In {\em Proceedings of the IEEE/CVF Conference on Computer Vision and
  Pattern Recognition (CVPR)}, June 2019.

\bibitem{Lu_2020_CVPR}
Yantao Lu, Yunhan Jia, Jianyu Wang, Bai Li, Weiheng Chai, Lawrence Carin, and
  Senem Velipasalar.
\newblock Enhancing cross-task black-box transferability of adversarial
  examples with dispersion reduction.
\newblock In {\em Proceedings of the IEEE/CVF Conference on Computer Vision and
  Pattern Recognition (CVPR)}, June 2020.

\bibitem{Zheng_2020_CVPR}
Haizhong Zheng, Ziqi Zhang, Juncheng Gu, Honglak Lee, and Atul Prakash.
\newblock Efficient adversarial training with transferable adversarial
  examples.
\newblock In {\em Proceedings of the IEEE/CVF Conference on Computer Vision and
  Pattern Recognition (CVPR)}, June 2020.

\bibitem{Inkawhich_2019_CVPR}
Nathan Inkawhich, Wei Wen, Hai~(Helen) Li, and Yiran Chen.
\newblock Feature space perturbations yield more transferable adversarial
  examples.
\newblock In {\em Proceedings of the IEEE/CVF Conference on Computer Vision and
  Pattern Recognition (CVPR)}, June 2019.

\bibitem{Latif:2018}
Siddique Latif, Rajib Rana, and Junaid Qadir.
\newblock Adversarial machine learning and speech emotion recognition:
  Utilizing generative adversarial networks for robustness.
\newblock {\em arXiv:1811.11402}, 2018.

\end{thebibliography}

\end{document}